# Predicted Brightness of Starlink Internet Satellites at 350 km

Anthony Mallama[1]

2024 June 24

[1] IAU - Centre for the Protection of Dark and Quiet
Skies from Satellite Constellation Interference

Correspondence: anthony.mallama@gmail.com

Abstract

SpaceX recently proposed to orbit 19,440 Starlink internet satellites at a low altitude of 350 km instead of the current 550 km. The distribution in the sky and the apparent magnitudes of these spacecraft are simulated in this paper. During astronomical twilight the impact of spacecraft at 350 km on astronomical observations would be more severe than those at 550 km. However, during the hours of darkness those at 350 km would have a less severe impact. The qualitative statement made by SpaceX to the US Federal Communications Commision is consistent with the quantitative results reported here.

1. Introduction

SpaceX recently proposed to orbit 19,440 Starlink internet satellites at 350 km instead of 550 km in order to reduce communication latency. In response, the US Federal Communications Commission sent the company a letter on June 7 asking for more details about the impact on optical astronomy. An excerpt from the reply SpaceX sent to the FCC is provided in Appendix A.

In this paper, we simulate the azimuth-elevation distribution of the proposed spacecraft at 350 km and estimate their apparent magnitudes in order to evaluate that impact quantitatively.

Mallama et al. (2024) recently reported on the observed brightness of Starlink Direct-to-Cell (DTC) satellites that are already orbiting at a similar altitude. These results are relevant because high atmospheric density at those low altitudes may require that spacecraft attitude is optimized to reduce drag rather than to fully implement brightness mitigation.

In fact, SpaceX recently informed us that the satellites we observed were not in brightness mitigation mode. This suggests that the brightness we determined for low altitude DTCs satellites is a worst case for predicting the luminosity of low altitude internet spacecraft. The best case is where the proposed satellites have the same luminosity characteristics as those that are brightness mitigated at 550 km.

2. Worst case scenario

We assume that brightness of the 19,440 internet satellites will correspond with the illumination phase function determined by Mallama et al. (2024) for low altitude DTCs. The illumination phase function for those satellites is

$$M = 7.719 - 0.0853\,\theta + 0.00115\,\theta^2 - 4.802\text{E-}6\,\theta^3$$

Equation 1

where $M$ is the visual magnitude adjusted to a distance of 1000 km and $\theta$ is the phase angle in degrees measured at the satellite between the directions to the Sun and the observer. Equation 1 is derived from a best fit solution over multiple spacecraft attitudes and observer-sun planes. Flares and attitude changes are not modeled, hence this formula predicts brightness in an average sense.

Simulated apparent magnitudes are computed from Equation 1 by supplying the phase angle and adjusting for distance. A further adjustment of 0.1 magnitude was added because the internet satellites are slightly smaller and thus likely fainter than DTCs. Finally, the dispersion of magnitudes around the best fitting phase function was applied using a pseudo-random number generator.

Satellite positions on the sky were determined by evaluating orbital parameters. The maps of internet satellites for solar elevations of -12º (the boundary between nautical and astronomical twilight), -18º (the boundary between astronomical twilight and darkness), and -24º are shown in Figures 1, 2 and 3, respectively. These are polar plots of the sky where the radial axis is elevation above the horizon and the polar angle is azimuth measured from north through east.

The magnitude distributions are listed in Table 1. Counts are cumulative, meaning that satellites brighter than mag 4 are also counted in those brighter than mag 5, etc. In all cases, the observer latitude is 30º and the time of year is equinox.

Table 1. Magnitude distributions from DTC phase function

```
Solar elev -->   -12      -18      -24
Brighter
than magn:
    4             58       33       13
    5            129       84       43
    6            229      185       99
    7            353      262      158
    8            461      331      203
    9            505      366      230
   10            519      377      242
```

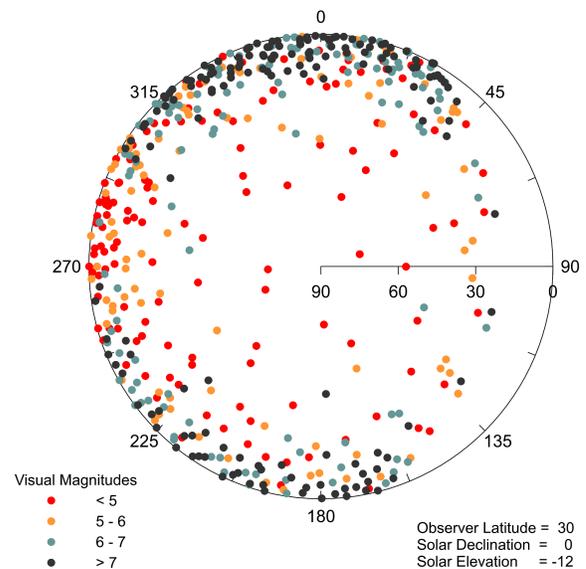

Figure 1. At solar elevation -12º many bright satellites are visible in nearly all parts of the sky. We assume that the orbits are evenly grouped in planes at 42º, 48º and 53º inclination based on previous orbits used by SpaceX.

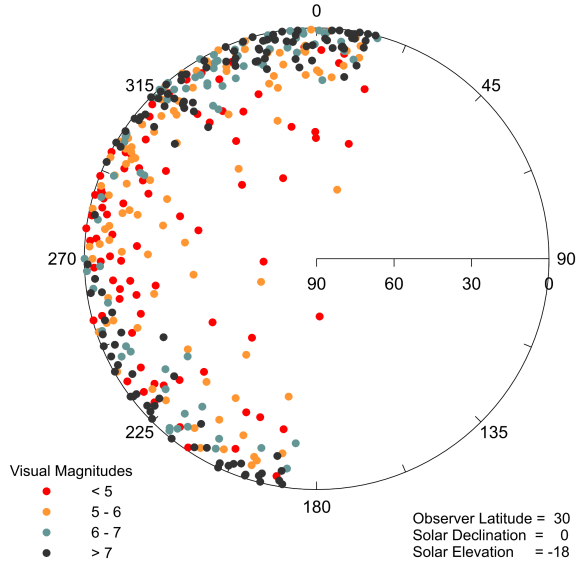

*Figure 2. At solar elevation -18° visible satellites are concentrated toward the solar azimuth of 281°. Those in the opposite direction are eclipsed by the Earth's shadow.*

Satellites brighter than magnitude 7 cause serious degradation of images from wide field astronomical instruments such as LSST (Tyson et al. 2020). Meanwhile those brighter than mag 6 are a distraction for amateur astronomers and naturalists. Table 1 indicates that at solar elevations of -12°, -18°, and -24°, the numbers of satellites brighter than mag 7 are 353, 262 and 158, respectively. The corresponding numbers of satellites brighter than mag 6 are 229, 185 and 99.

3. Best case scenario

We assume that brightness mitigation will be fully optimized. So, low altitude internet satellites will have the same brightness characteristics as those at high altitudes.

The phase function for internet satellites is

$M = 5.822 - 0.00879\,\theta + 0.000848\,\theta^2 - 5.784\text{E-}6\,\theta^3$

Equation 2

Satellite positions are identical to those in Figures 1, 2 and 3, however the spacecraft are less luminous as shown in Table 2. At solar elevations of -12°, -18°, and -24°, the numbers of satellites brighter than magnitude 7 are 232, 174, and 130, respectively. The corresponding numbers of satellites brighter than mag 6 are 156, 108 and 71.

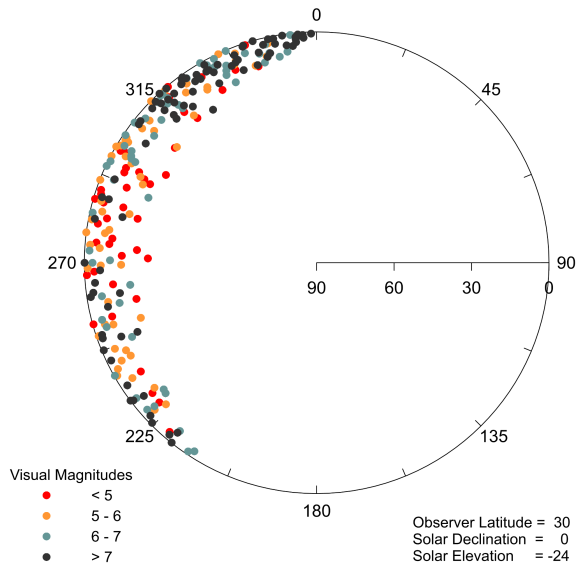

*Figure 3. At solar elevation -24° satellites are only visible low in the sky toward the solar azimuth of 285°.*

Table 2. Magnitude distributions
from internet phase function

```
Solar elev -->  -12     -18     -24
Brighter
than magn:
    4            41      15       0
    5            87      55      21
    6           156     108      71
    7           232     174     130
    8           335     263     182
    9           459     351     231
   10           511     385     244
```

4. Discussion

During astronomical twilight the number of Starlink internet satellites at 350 km brighter than magnitude 7 will be between 174 and 353 with an average value of 263. The number brighter than mag 6 will be between 108 and 229 with an average of 168.

For comparison purposes, the magnitudes of 19,440 satellites at 550 km were computed for the twilight condition using Equation 2. The average number brighter than mag 7 is 240, while that average for mag 6 is 126. These values are less than those quoted above for satellites at 350 km. So, the impact on astronomy during astronomical twilight is more severe for the 350 km altitude.

However, during darkness when the solar elevation is -24º, the counts of magnitudes brighter than 7 are the same for satellites at 350 and 550 km (both 144). Meanwhile, the count for mag 6 is less for satellites at 350 km (39) as compared to those at 550 km (85). Furthermore, the satellites at 350 km are only visible near the horizon toward the solar azimuth as shown in Figure 3 while those at 550 km extend nearly to zenith as illustrated in Figure 4.

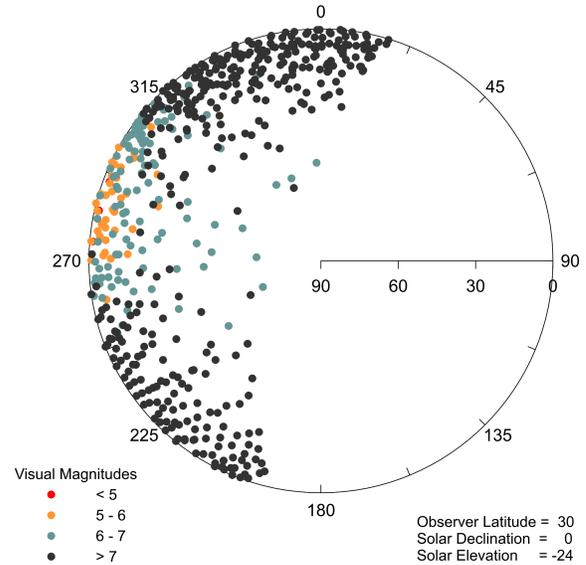

Figure 4. Distribution and magnitude for satellite at 550 km with solar elevation -24º.

These results apply to an observer at 30º latitude, which is near to that of many major astronomical observatories. The equinox time is a mean value for the whole year.

5. Conclusion

The impact on optical astronomy of 19,440 Starlink internet satellites at 350 km has been assessed. Worst case and best case scenarios were evaluated.

Spacecraft at 350 km would have a more negative impact than those at 550 km during astronomical twilight. However, during hours of darkness satellites at 350 km would have a less severe impact.

The qualitative statement made by SpaceX to the US Federal Communications Commision (Appendix A) is consistent with the quantitative results in this paper.

Acknowledgment

Richard E. Cole provided insight into the relationships between atmospheric drag, spacecraft attitude and brightness. An anonymous reviewer from the IAU-CPS provided important comments that improved this paper.

Appendix A.

Excerpt from SpaceX reply to the US FCC dated 2024 June 20.

"SpaceX continues to work closely with NSF to ensure any increased brightness from satellites operating at lower altitudes is offset by the benefits of transiting more quickly through the field of view, being visible to fewer observers, and passing into Earth's shadow sooner after sunset. SpaceX will continue to work with NSF to develop mitigation solutions and coordinate operations."

Appendix B.

Excerpt from the abstract of Mallama et al. (2024).

"The mean apparent magnitude of Starlink Mini Direct-To-Cell (DTC) satellites is 4.62 while the mean of magnitudes adjusted to a uniform distance of 1000 km is 5.50. DTCs average 4.9 times brighter than other Starlink Mini spacecraft at a common distance."